\documentclass[a4paper,fleqn,usenatbib]{mnras}

\usepackage{newtxtext,newtxmath}

\usepackage[T1]{fontenc}
\usepackage{ae,aecompl}

\usepackage{graphicx}	
\usepackage{amsmath}	
\usepackage{amssymb}	

\newcommand{\dem}{DEM~L241}
\newcommand{\cxou}{CXOU~053600.0$-$673507}

\newcommand{\pthree}{LMC~P3}

\newcommand{\psrj}{PSR\,J2032$+$4127}
\newcommand{\ls}{LS~5039}

\newcommand{\psrb}{PSR~B1259$-$63/LS~2883}

\newcommand{\fermi}{$Fermi$-LAT}

\title[The orbital parameters of \pthree]{The orbital parameters of the gamma-ray binary \pthree\thanks{Based on observations made with the Southern African Large Telescope (SALT) under program 2016-1-MLT-006 (PI: N. Komin).}}

\author[B. van Soelen, N. Komin, A. Kniazev, P. V\"ais\"anen]{
B. van Soelen,$^{1}$\thanks{E-mail: vansoelenb@ufs.ac.za}
N. Komin,$^{2}$\thanks{E-mail: nukri.komin@wits.ac.za}
A. Kniazev$^{3,4}$, P. V\"ais\"anen$^{3,4}$ 
\\
$^{1}$Department of Physics, University of the Free State,  PO Box 339, Bloemfontein 9300, South Africa\\
$^{2}$School of Physics, University of the Witwatersrand, 1 Jan Smuts Avenue, Braamfontein, Johannesburg, 2050 South Africa\\
$^3$South African Astronomical Observatory, PO Box 9, Observatory, 7935,  Cape Town, South Africa \\
$^4$Southern African Large Telescope,  PO Box 9, Observatory, 7935, Cape Town, South Africa
}

\date{Accepted XXX. Received YYY; in original form ZZZ}

\pubyear{2017}

\begin{document}
\label{firstpage}
\pagerange{\pageref{firstpage}--\pageref{lastpage}}
\maketitle

\begin{abstract}

\pthree\ is the most luminous gamma-ray binary discovered to date and the first detected outside of the Galaxy, with an orbital period of $10.301$\,d. We report on optical spectroscopic observations undertaken with the Southern African Large Telescope (SALT) using the High Resolution spectrograph (HRS). We find the binary is slightly eccentric, $e = 0.40\pm0.07$, and place the time of periastron at HJD $2457412.13 \pm 0.29$. Stellar model fitting finds an effective temperature of $T_\rmn{eff} = 36351 \pm 53$\,K. The mass function, $f =  0.0010 \pm 0.0004$\,M$_{\sun}$, favours a neutron star compact object. The phases of superior and inferior conjunctions are $0.98$ and $0.24$, respectively (where phase 0 is at the {\it Fermi}-LAT maximum), close to the reported maxima in the GeV and TeV light curves.

\end{abstract}




\section{Introduction}

Gamma-ray binaries are a distinct class of high mass binary systems, defined by having spectral energy distributions that peak (in a $\nu F_\nu$ distribution) in the gamma-ray regime \citep[see e.g.][for a detailed review of these sources]{2013A&ARv..21...64D}. There are only seven such systems  known, the most recent of which, \psrj, was recently detected at very high energies around periastron in November 2017 \citep{ho17,mirzoyan17,veritas_magic_j2032_2018}. In this paper we present high resolution spectroscopic optical observations of the recently discovered source \pthree, the most luminous of the gamma-ray binaries and the first detected outside of the Galaxy, lying in the Large Magellanic Cloud \citep[LMC;][]{Fermi}.

All gamma-ray binaries consist of a compact object, within the mass range of a neutron star or a black hole, which is in orbit around an O or B type star \citep{2013A&ARv..21...64D,Fermi}. However, for only two systems, \psrb\ and the recently detected \psrj, is the nature of the compact known, since they have been detected as pulsars \citep{1992ApJ...387L..37J, abdo09}. In these systems, the non-thermal emission is believed to arise due to the particle acceleration that occurs at the shock that forms between the pulsar and stellar winds. In the other sources a black hole compact object cannot be ruled out, and microquasar scenarios are still considered.

\pthree\ was a point-like source ``P3''  detected in \fermi\ observations of the LMC \citep{FermiLMC}. The binary nature of \pthree\ was discovered through a search for periodicity in \fermi\ observations, finding a $10.301\pm0.002$\,d period \citep{Fermi}. \pthree\ is associated with the previously detected point-like X-ray source \cxou\ located in the supernova remnant \dem\ \citep{Bamba}. It was previously suggested by \citet{Seward} that based on the variability of the X-ray flux and small variations of the radial velocity of an O5III(f) star (V = 13.5) coincident with this X-ray source, that the object was a High-Mass X-ray Binary (HMXB) with a period of tens of days. 
X-ray and radio observations confirmed the multi-wavelength modulation on the same 10.301\,d period while optical radial velocity measurements of the O5III(f) star (the earliest type of any gamma-ray binary) also showed a variation consistent with this period \citep{Fermi}. 
The binary solution to the radial velocities found a mass function of $ f(M) = \left(1.3^{+1.1}_{-0.6} \right) \times 10^{-3}$\,M$_\odot$, however, the eccentricity of the system could not be constrained.

The radio and X-ray light curves are in phase, but are in anti-phase with the \fermi\ light curves \citep{Fermi}. The H.E.S.S.\ telescope has subsequently reported detection at TeV energies (though only in a single phase bin) which is also in anti-phase with the \fermi\ observations \citep{hess17_binary}. This is very similar to what is observed for the gamma-ray binary \ls\ \citep{aharonian05,kishishita09,abdo09}.  Key to understanding the gamma-ray emission is obtaining a clear solution for the binary parameters of the source. Here we report on optical spectroscopic observations of \pthree\ undertaken with the Southern African Large Telescope (SALT) using the High Resolution spectrograph (HRS), to establish the binary parameters.

\section{Observations}

\subsection{Observations and data reduction}

\pthree\ was successfully observed 24 times with the High Resolution Spectrograph \citep[HRS;][]{bramall10,bramall12,crause14} on the Southern African Large Telescope \citep[SALT;][]{buckley_06} between 2016 September 14 and 2017 February 06. 

The HRS is a dual-beam fibre-fed \'echelle spectrograph, housed within a vacuum tank, inside a thermo-stable room. The HRS is designed for extra-solar planet searches with velocity accuracies of 5\,m\,s$^{-1}$ in the High Stability Mode. As part of the HRS calibration plan, flats and ThAr hollow-cathode lamp spectra are obtained weekly through both the object and sky fibres. Observations of radial velocity standards are taken as part of the HRS calibration plan.\footnote{The full details of the calibration plan and the stability of the radial velocitiy determinations are given in the SALT proposal call documentation { http://pysalt.salt.ac.za/proposal\_calls/current/ProposalCall.html}. }

Observations were undertaken using the Low Resolution Mode ($R=14\,000$), with each observation consisting of two camera exposures of 1\,220~s, (except for two nights where the exposure was increased to 2$\times$1\,640~s). The different orders of the HRS spectra were extracted and wavelength calibrated using the HRS pipeline discussed in \citet{kniazev16}. Each individual order of the spectrum was normalized and merged into a single one dimension spectrum using the standard {\sc iraf}/{\sc p}y{\sc raf} packages. Heliocentric correction was performed for each individual exposure using {\sc rvcorrect/dopcor} and then nightly observations were averaged together. 

HRS observations are undertaken with a 2.2\arcsec\ fibre placed on the target and a separate ``sky fibre'' that must be placed at least 16\arcsec\ away from the target.  Because the target lies within a nebula the sky lines are dominated by the Balmer emission lines arising from the nebula, and the background sky measured in the sky fibre was significantly different from the sky as measured at the target. As a result, the sky subtraction was not able to properly correct for the nebula emission and introduced more noise into the spectrum. For this reason, no sky subtraction was performed and the analysis was restricted to the ``blue'' arm of the HRS where the nebula and sky contamination was minimal.

\subsection{Radial velocity determination}

The radial velocity was investigated by fitting the position of individual lines and by cross-correlating the spectrum to a template. 

The observed central wavelength of different absorption lines was determined by fitting Gaussian profiles, which showed that different line species have different radial velocities. This effect has previously been noted in O-type stars, and is most likely due to the contamination by the stellar wind \citep[see e.g.][]{casares05,sarty11,puls96,waisberg15}.

\begin{figure}
 \includegraphics[width=\columnwidth]{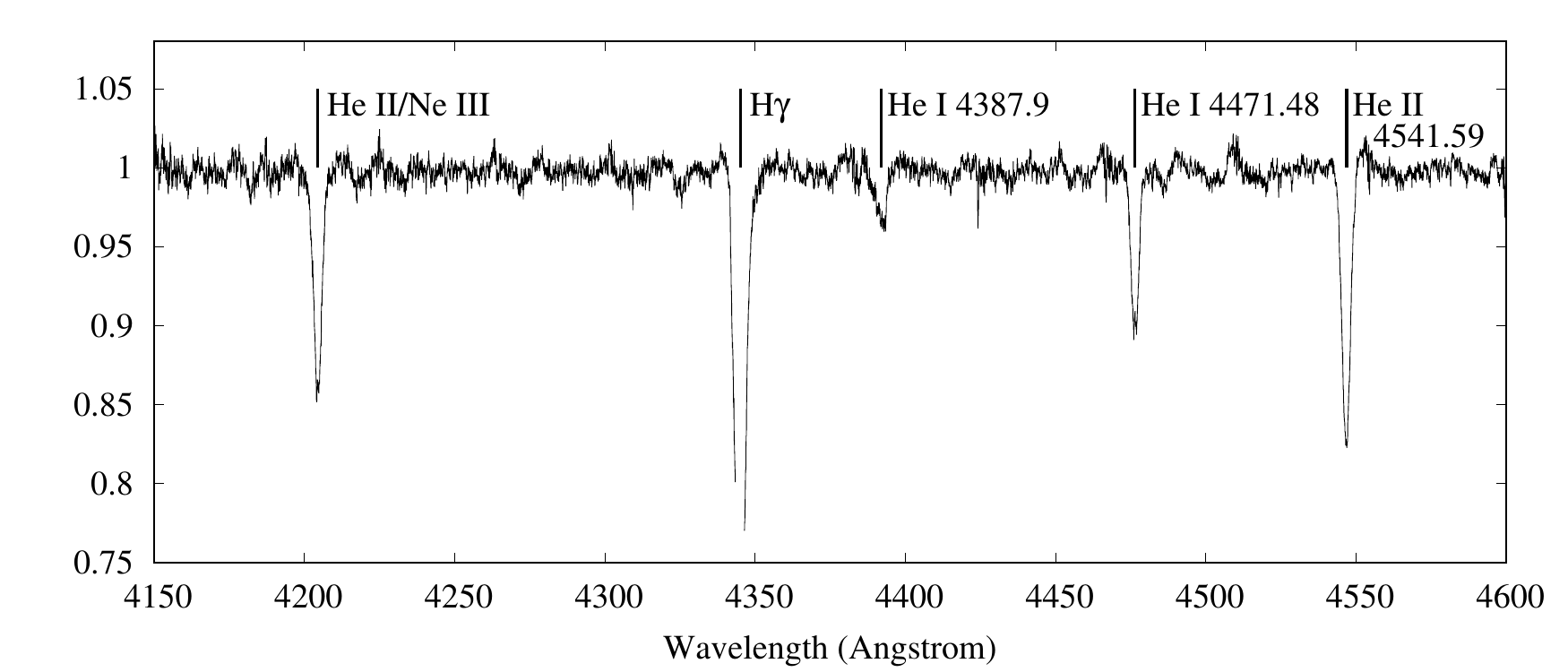}
 \caption{Template spectrum created from averaging over all observations.  The gap in the H$\gamma$ line  is where a section was excluded from the cross-correlation calculation because of contamination by a emission line from the nebula.}
 \label{fig:template_spectrum}
\end{figure}

Because of the different velocities found for different lines the radial velocity was determined by cross-correlating individual spectra to a template, using the {\sc rvsao/xcsao} package \citep{kurtz98}. We followed a similar process to that described in, for example, \citet{manick15,foellmi03,monageng17}, and created the template from the available observations. 
To create the reference template, first an average of all observations was found and the velocity shift between each observation and the average spectrum was determined through cross-correlation. Next, all the individual spectra were corrected by the shift to the average spectrum and the template was produced by averaging these velocity corrected spectra. The final template spectrum, in the 4150--4600\,\AA{} wavelength range  used for the cross-correlation, is shown in Fig.~\ref{fig:template_spectrum}.

In order to determine the zero-velocity of the template spectrum, we used the ULySS program \citep{koleva09} with a medium spectral-resolution MILES
library  to  simultaneously determine the line-of-sight velocity for the star, and its T$_\rmn{eff}$, log~g and [Fe/H]. A fit over the 4160--5000\,\AA{} wavelength range finds a redshift of 
$cz = 320.7 \pm 0.7$\,km~s$^{-1}$ (with a dispersion of $58.9 \pm 1.2$\,km~s$^{-1}$). 
The stellar model fitting method also provides a best fit to the atmospheric properties of the star, finding an effective temperature of 
$T_\rmn{eff} = 36351 \pm 53$\,K, a surface gravity $\log g = 3.4 \pm 0.1$\,$[\log($cm~s$^{-2})]$, and a metallicity of $[\rmn{Fe/H}] = 0.25 \pm  0.01$.

This is compatible with an OIII type star, though the values are lower than those of an O5 III star in, for example, \citet{martins05}. However, the exclusion of parts of the Balmer lines prevents us from undertaking more detailed stellar atmospheric modelling.

The final radial velocity, relative to the template, was calculated by performing the cross-correlation analysis in the 4150--4600\,\AA{} wavelength range which contains four He lines and the H$\gamma$ line, and which has limited contamination from the sky lines and the nebula. The wavelength region 4343.5--4346.5\,\AA{} was ignored in the cross-correlation analysis as it contained  significant contamination from a narrow nebula emission line superimposed on the stellar absorption line. 

\subsection{Orbital parameter determination}

The binary orbital parameters were determined from the fit to the radial velocities using the {\sc{helio\_rv}} package which is part of the IDL Astronomy Library \citep{landsman93}.\footnote{https://idlastro.gsfc.nasa.gov/} All reported errors on the fitted binary parameters are calculated by scaling the errors in the radial velocity measurements to achieve a  reduced $\chi^2$ of exactly 1 \citep{lampton76}. This scaling factor was $\sim1.2 - 5.2$ depending on the data.
 This does not change the values of the fitted parameters, but gives a more accurate estimate of the error, since addition systematic errors are better accounted for.

Table~\ref{tab:orb_line_fit} shows the orbital parameters determined from the velocities of the individual lines. All fits were performed assuming a fixed period of 10.301~d.
We find different velocities from the He I and He II lines, as was noted above. The measurements of the individual lines do suffer from lower signal-to-noise and for H$_\beta$ contamination from a nebula emission line.

The solution using the radial velocities determined from cross-correlation is shown in Table~\ref{tab:orb_cross}. If the orbital period is kept as a free parameter, the best-fitting orbital period is $10.314 \pm 0.044$\,days, which is consistent with the $10.301 \pm 0.002$\,d period found from the \fermi\ data \citep{Fermi}. However, searches for periodicity using the Lomb-Scargle technique could not detect a statistically significant period and we have, therefore, adopted an orbital period of 10.301\,days for our final result. 

\begin{figure}
 \includegraphics[width=\columnwidth]{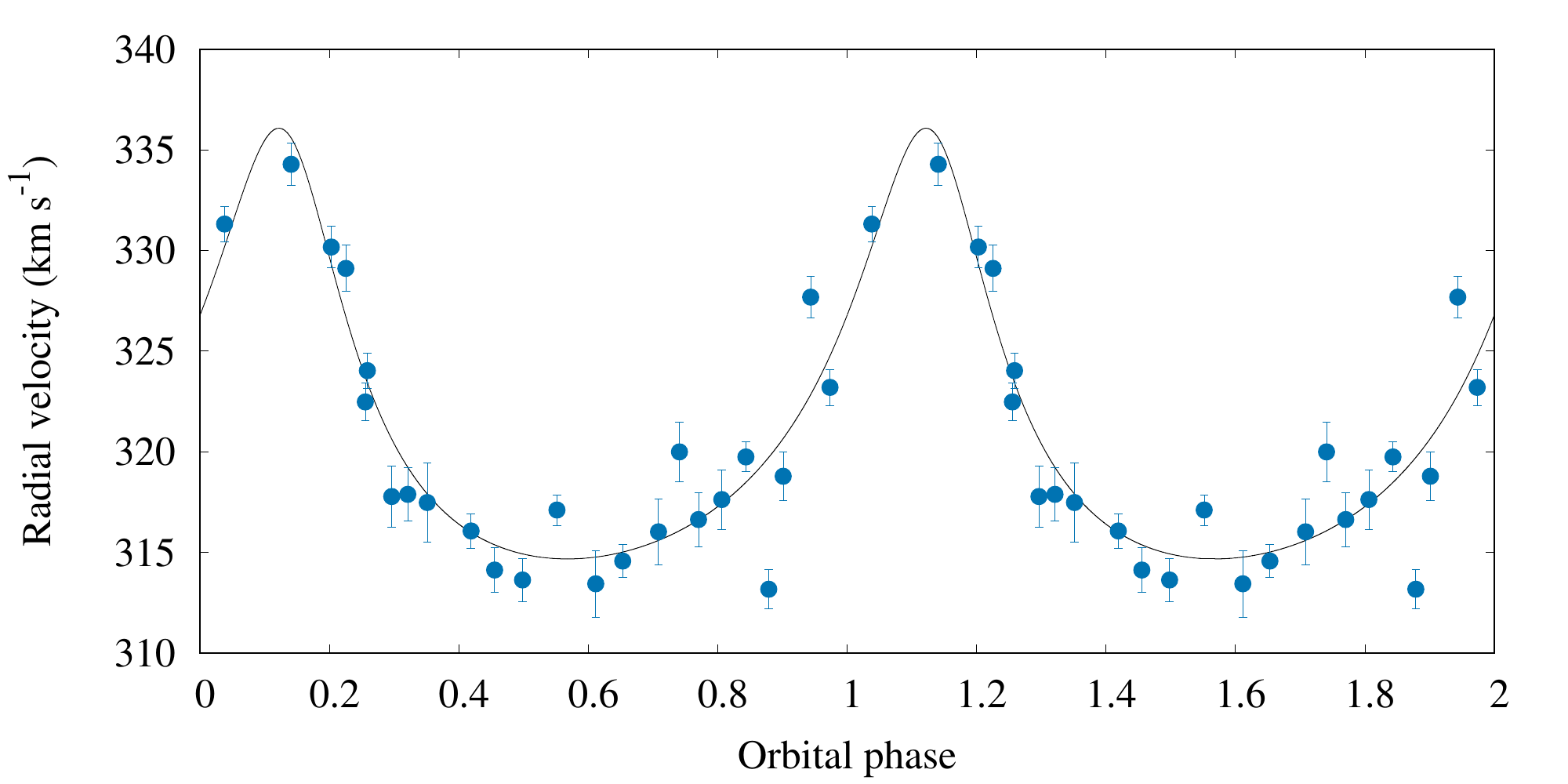}
 \caption{Radial velocity of the O5III star in the binary system, as determined from cross-correlation. The solid line shows the best fit to this data, with a fixed 10.301\,day orbital period. For clarity the plot is reported over two orbital phases with phase $\phi=0$ at MJD = 57\,410.25, which corresponds to the given phase in \citet{Fermi}. The error bars show the statistical errors reported by the {\sc rvsao} package. }
 \label{fig:rv_4160_4600}
\end{figure}

We do note the systemic velocity we find, $\Gamma = 321.18 \pm 0.85$\,km~s$^{-1}$, is higher than the $295.8\pm2.0$\,km~s$^{-1}$ previously reported \citep{Fermi}. We undertook additional analysis to confirm that the wavelength calibration performed by the HRS pipeline was correct and that the comparison to the radial velocity standards was accurate to within the expected performance. We found no evidence of any discrepancy in the calibration, nor any long term systematic shift in the wavelength calibration over the period of observations.
We believe that there are two possible reasons for this difference; there may be a possible systematic offset arising from the fit of the high resolution template fit to the medium resolution MILES libraries (however there is no significant offset to the radial velocity standards) or there may have been a systematic offset in the zero velocity in the field O-type star used in the previous analysis. However, this difference does not change the main results of determining the orbital parameters. If this difference in the systemic velocity is removed the radial velocities are in agreement with the previous results.
The final radial velocity curve and the best-fitting model are shown in Fig.~\ref{fig:rv_4160_4600}, and the data are given in Table~\ref{tab:rv_values}. We find the binary is slightly eccentric, $e = 0.40\pm0.07$, and place the time of periastron at HJD $2457412.13 \pm 0.29$.

\begin{table*}
\caption{Orbital parameters as determined from the radial velocities calculated by Gaussian fits to individual lines.}
\label{tab:orb_line_fit}
 \begin{tabular}{lccccc} \hline
 &  H beta   & He I 4471   & He I 4921   & He II 4541   & He II 5411   \\ \hline
Time of periastron (HJD) &$ 2457412.15 \pm 0.50 $&$ 2457412.45 \pm 0.21 $&$ 2457411.89 \pm 0.36 $&$ 2457411.99 \pm 0.21 $&$ 2457412.14 \pm 0.33 $ \\
Orbital period (fixed, days) &$ 10.301 \pm 0.000 $&$ 10.301 \pm 0.000 $&$ 10.301 \pm 0.000 $&$ 10.301 \pm 0.000 $&$ 10.301 \pm 0.000 $ \\
Systemic velocity (km/s) &$ 298.43 \pm 1.26 $&$ 331.02 \pm 0.75 $&$ 335.95 \pm 2.77 $&$ 340.00 \pm 0.48 $&$ 337.33 \pm 0.71 $ \\
K (velocity semi-amplitude) &$ 13.80 \pm 2.84 $&$ 15.14 \pm 2.10 $&$ 24.23 \pm 26.31 $&$ 11.30 \pm 1.21 $&$ 11.87 \pm 1.68 $ \\
Eccentricity &$ 0.40 \pm 0.12 $&$ 0.52 \pm 0.07 $&$ 0.69 \pm 0.28 $&$ 0.42 \pm 0.06 $&$ 0.46 \pm 0.08 $ \\
Longitude of periastron (degrees) &$ 5.9 \pm 20.9 $&$ 29.9 \pm 10.4 $&$ 354.5 \pm 10.9 $&$ 12.0 \pm 8.9 $&$ 8.1 \pm 13.34 $ \\
Mass function (M$_{\sun}$) &$ 0.0021 \pm 0.0014 $&$ 0.0023 \pm 0.0010 $&$ 0.0058 \pm 0.0201 $&$ 0.0011 \pm 0.0004 $&$ 0.0013 \pm 0.0006 $ \\ \hline
 \end{tabular}

\end{table*}

\begin{table*}
\caption{Orbital parameters determined from the velocities calculated by cross-correlation. Note the systemic velocity is relative to the template file used. }
\label{tab:orb_cross}
 \begin{tabular}{lcc} \hline
 Parameters & Free & Fixed \\
 & & (adopted) \\ \hline
Time of periastron (HJD) &$ 2457411.77 \pm 1.34 $&$ 2457412.13 \pm 0.29 $ \\
Orbital period ( days) &$ 10.314 \pm 0.044 $&$ 10.301 \pm 0.000 $ \\
Systemic velocity relative to template (km/s) &$ 0.73 \pm 0.59 $&$ 0.68 \pm 0.55 $ \\
Systemic velocity (km/s) &$ 321.23 \pm 0.88 $&$ 321.18 \pm 0.85 $ \\
K (velocity semi-amplitude)&$ 10.69 \pm 1.24 $&$ 10.69 \pm 1.23 $ \\
Eccentricity &$ 0.39 \pm 0.08 $&$ 0.40 \pm 0.07 $ \\
Longitude of periastron (degrees) &$ 12.9 \pm 12.8 $&$ 11.3 \pm 12.0 $ \\
Mass function (M$_{\sun}$) &$ 0.0010 \pm 0.0004 $&$ 0.0010 \pm 0.0004 $ \\ \hline

 \end{tabular}

\end{table*}

\begin{table}
 \caption{Radial velocities obtained from cross-correlation}
 \label{tab:rv_values}
 
 \begin{tabular}{lc} \hline
  HJD & Radial velocity (km~s$^{-1}$)    \\ \hline
2457655.60791 & $ 316.64 \pm 1.35 $ \\
2457665.60399 & $ 320.00 \pm 1.48 $ \\
2457691.50654 & $ 322.48 \pm 0.94 $ \\
2457708.46002 & $ 318.79 \pm 1.20 $ \\
2457711.56664 & $ 330.17 \pm 1.03 $ \\
2457722.44354 & $ 324.03 \pm 0.87 $ \\
2457723.39659 & $ 317.48 \pm 1.96 $ \\
2457724.46868 & $ 314.13 \pm 1.11 $ \\
2457725.46161 & $ 317.11 \pm 0.76 $ \\
2457728.46743 & $ 319.75 \pm 0.74 $ \\
2457730.47147 & $ 331.31 \pm 0.87 $ \\
2457731.53120 & $ 334.28 \pm 1.04 $ \\
2457732.40171 & $ 329.11 \pm 1.15 $ \\
2457733.38795 & $ 317.89 \pm 1.32 $ \\
2457734.39469 & $ 316.07 \pm 0.86 $ \\
2457736.37916 & $ 313.45 \pm 1.65 $ \\
2457737.37704 & $ 316.03 \pm 1.64 $ \\
2457738.38493 & $ 317.63 \pm 1.48 $ \\
2457774.33596 & $ 317.78 \pm 1.51 $ \\
2457776.41725 & $ 313.64 \pm 1.06 $ \\
2457780.33657 & $ 313.18 \pm 0.99 $ \\
2457781.31273 & $ 323.20 \pm 0.89 $ \\
2457788.31389 & $ 314.58 \pm 0.82 $ \\
2457791.30522 & $ 327.68 \pm 1.04 $ \\ \hline
 \end{tabular}

\end{table}

\section{Discussion}

\subsection{Mass of the compact object}

\begin{figure}
 \includegraphics[width=\columnwidth]{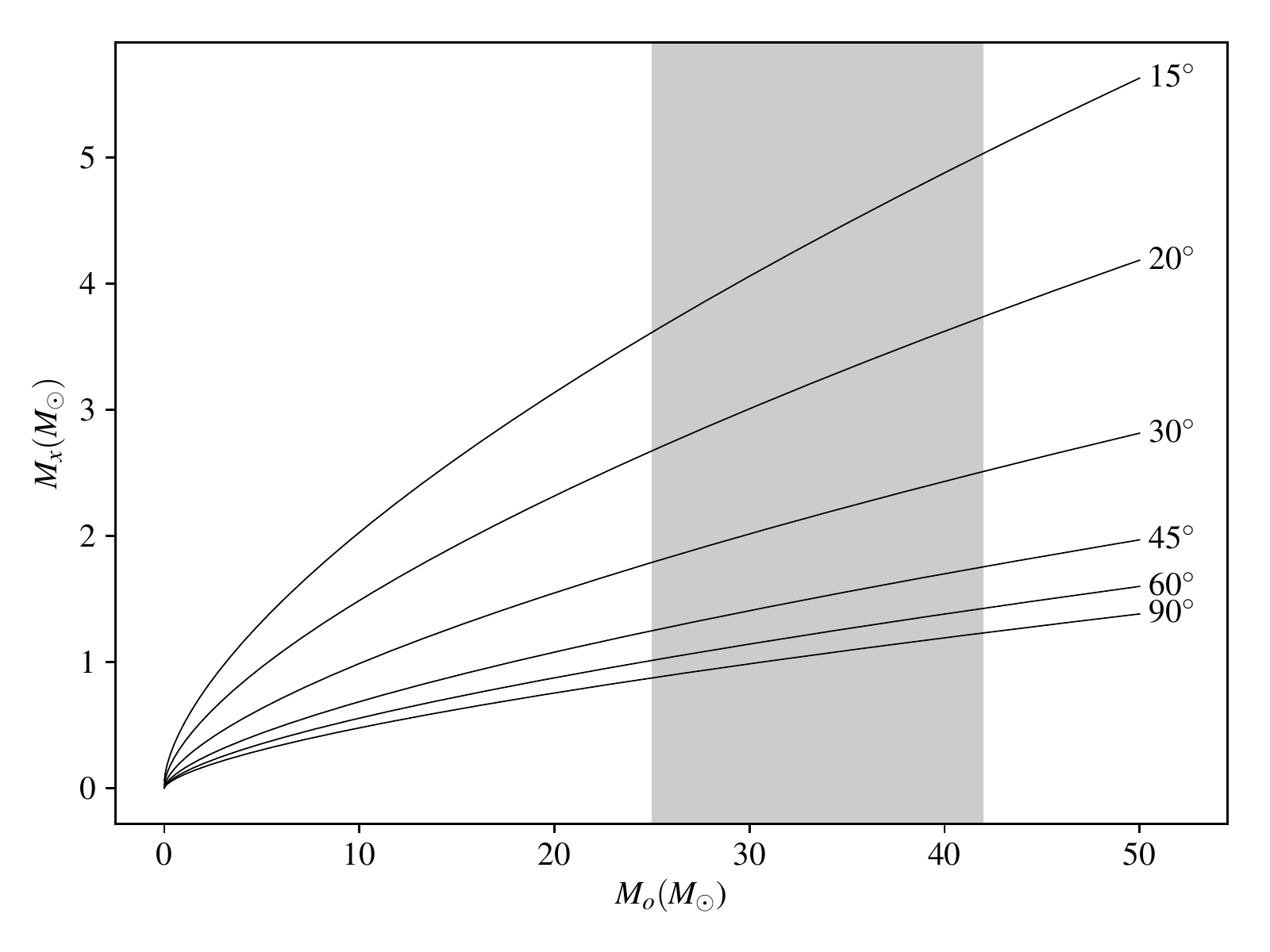}
 \caption{The constraint on the mass of the compact object in \pthree. The shaded area marks the range of assumed masses of the optical companion.}
 \label{fig:mass_function}
\end{figure}

The binary parameter solution gives a mass function of $f =  0.0010 \pm 0.0004$\,M$_{\sun}$ and the constraints on the mass of the compact object are shown in Fig.~\ref{fig:mass_function}. The mass of an O5III(f) star is $\sim40$\,M$_{\sun}$ \citep{martins05}, but due to the binary evolution could be different.  A mass range of $25 - 42$M$_{\sun}$ was considered in \citet{Seward}. For this mass range both a neutron star or black hole mass is compatible with the mass function, though a neutron star is favoured.
For a $40$\,M$_{\sun}$ star, the mass of the compact object will be $>5$\,M$_{\sun}$ for inclinations $i \leq 15 \pm 2$\degr, and  while the inclination must be $i\leq 11 \pm 1$\degr for a $25$\,M$_{\sun}$  optical companion.
 Assuming the compact object is a pulsar, with a mass of  $1.4$\,M$_{\sun}$, the inclination will lie between $i=39\pm6$\degr and $i = 59\pm11$\degr for a $25$\,M$_{\sun}$ and $40$\,M$_{\sun}$ star, respectively.

\subsection{Binary orientation}

\begin{figure}
 \includegraphics[width=\columnwidth]{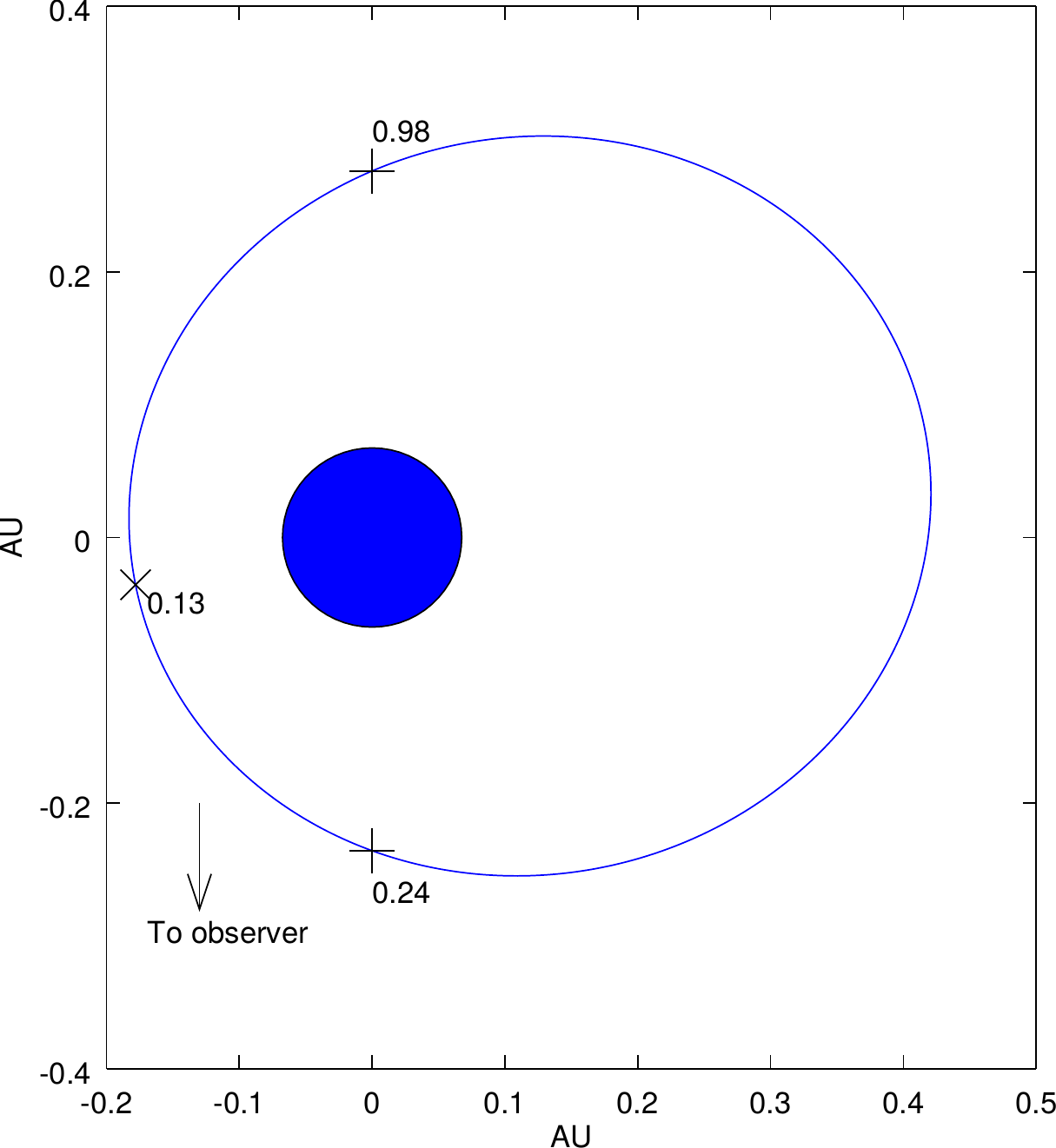}
 \caption{The binary orientation of the LMC~P3. This is calculated assuming a $M_\star = 33.5$\,M$_\rmn{\sun}$ and radius of $R_\star = 14.5$\,R$_\rmn{\sun}$ and $M_p = 1.4$\,M$_\rmn{\sun}$. The blue circle shows the relative size of the optical star while the black line traces the orbit of the compact object. The positions and orbital phase of superior and inferior conjunction are marked by $+$ while the position and phase of periastron is marked by a $\times$.}
 \label{fig:binary_parameters}
\end{figure}

The orientation of the binary system is shown in Fig.~\ref{fig:binary_parameters}. The system parameters are calculated assuming the optical star has a mass of $M_\star = 33.5$\,M$_\rmn{\sun}$ and a radius of $R_\star = 14.5$\,R$_\rmn{\sun}$  \citep[average of the reported mass range and the corresponding average radius; e.g.][]{martins05}	  
and a mass of $M_p = 1.4$\,M$_\rmn{\sun}$ for the compact object. 
 This would correspond to an inclination of $i\approx50$\degr.
Following \citet{Fermi} phase $\phi =0$  is assumed to be at MJD = 57410.25.
Superior conjunction (when the compact object is behind the optical star) occurs at $\phi_\rmn{sup} = 0.98$ and inferior conjunction is at $\phi_\rmn{inf} = 0.24$ with periastron occurring at $\phi_\rmn{per} = 0.13$.

\subsection{Implications for gamma-ray emission}

The SALT HRS observations have shown that LMC~P3 has an eccentricity of $\approx 0.4$ and established the orientation (longitude of periastron) of the system. Both the eccentricity and the orientation play an important role in the modulation of the observed gamma-ray emission, since the inverse Compton scattering is dependent on the energy density of the target photons and the angle of scattering.  In the cases of the highly eccentric systems \psrb\ and \psrj\ gamma-ray emission is only detected near periastron \citep[e.g.][]{hess2013,veritas_magic_j2032_2018}.

The \fermi\ observations of \pthree\ show a flux, peaking slightly after phase $\phi \sim 0$ \citep{Fermi} while the H.E.S.S.\ telescope has only detected the VHE emission, in a single phase bin of $\phi_{\rm bin} = 0.2-0.4$ \citep{hess17_binary}.  Such an out of phase light curve between the GeV and TeV gamma-ray emission can arise since  $\gamma\gamma$ absorption can modulate very high energy emission \citep[e.g.][]{2006A&A...451....9D,boettcher05}. In this scenario, because of the strong angular dependence of inverse Compton scattering and $\gamma \gamma$ absorption,  both the maximum in the inverse Compton emission and $\gamma \gamma$ absorption should occur around superior conjunction. This could lead to a maximum in the GeV light curve near superior conjunction since the photons are below the pair-production threshold energy, while the maximum in the TeV light curve will occur around inferior conjunction  where the $\gamma\gamma$ opacity is lowest. The binary solution found in this paper supports this for the \pthree,  with superior conjunction lying at $\phi\approx0.98$ at the peak of the \fermi\ light curve, while  inferior conjunction, $\phi\approx0.24$ is around the peak in the H.E.S.S.\ light curve.

Additionally Doppler boosting of the emission may play a role. The tail of the shock may obtain a relativistic bulk velocity, as is evident from hydrodynamical simulations of gamma-ray binaries \citep[e.g.][]{bogovalov08,bogovalov12}. If, due to $\gamma\gamma$ absorption, we predominately observe TeV gamma-ray emission originating from this region (with the GeV emission originating from the apex of the shock), this would lead to an enhancement in the observed TeV emission near inferior conjunction when the material is directed towards us \citep[see e.g.][]{dubus10,zabalza13}.

\section{Conclusion}

We have undertaken  SALT/HRS observations of \pthree\ and established the best orbital parameters for this system so far. The best-fitting solution shows the binary has an eccentricity of $e = 0.40\pm0.07$, which makes it similar to \ls.  The best orbital parameter fit places superior conjunction at orbital phase $\phi = 0.98$, close to the maximum in the \fermi\ light curve, while inferior conjunction is at phase $\phi=0.24$. This orientation may explain the anti-phase between the GeV and TeV light curves. The determined mass function, $f = 0.0010\pm 0.0004$\,M$_{\sun}$, favours a neutron star compact object, and subsequently favours a pulsar wind driven and not accretion driven system.

\section*{Acknowledgements}

The authors are grateful to P.A.\ Charles, A.\ Odendaal, A.F.\ Rajoelimanana and L.J.\ Townsend for valuable discussions. 
All of the observations reported in this paper were obtained with the Southern African Large Telescope (SALT). 
BvS and NK acknowledge that this work was supported by the Department of Science and
Technology and the National Research Foundation of South Africa through a block grant to the South African Gamma-Ray
Astronomy Consortium. AK and PV acknowledge the support of the National Research Foundation of South Africa.




\bibliographystyle{mnras}
\bibliography{LMC_P3} 






\bsp	
\label{lastpage}
\end{document}